\documentclass[aps,11pt,prd,groupedaddress,nofootinbib,notitlepage,eqsecnum,preprintnumbers]{revtex4-2}
\usepackage[utf8]{inputenc}
\usepackage{hyperref}
\usepackage{xcolor}
\usepackage{graphicx}
\usepackage{amsmath,amssymb}
\usepackage{bm}
\usepackage{comment}
\usepackage[shortlabels]{enumitem}
\usepackage{overpic}
\usepackage{ulem}
\usepackage{makecell}
\usepackage{tabularx}
\usepackage{booktabs}
\usepackage{diagbox}
\usepackage{pifont}
\usepackage{orcidlink}

\DeclareMathOperator{\arccosh}{arccosh}

\begin{document}
\title{Finite distance effects on the Hellings-Downs curve in modified gravity}

\author{\textsc{Guillem Dom\`enech$^{a,b}$\,\orcidlink{0000-0003-2788-884X}}}
    \email{{guillem.domenech}@{itp.uni-hannover.de}}
\author{\textsc{Apostolos Tsabodimos$^{a}$\,\orcidlink{0009-0001-0230-5647}}}
    \email{{apostolos.tsabodimos}@{stud.uni-hannover.de}}

\affiliation{$^a$Institute for Theoretical Physics, Leibniz University Hannover, Appelstraße 2, 30167 Hannover, Germany.}
\affiliation{$^b$ Max-Planck-Institut für Gravitationsphysik, Albert-Einstein-Institut, 30167 Hannover, Germany}

\begin{abstract}
There is growing interest in the overlap reduction function in pulsar timing array observations as a probe of modified gravity. However, current approximations to the Hellings-Downs curve for subluminal gravitational wave propagation, say $v<1$, diverge at small angular pulsar separation. In this paper, we find that the overlap reduction function for the $v<1$ case is sensitive to finite distance effects. First, we show that finite distance effects introduce an effective cut-off in the spherical harmonics decomposition at $\ell\sim \sqrt{1-v^2} \, kL$, where $\ell$ is the multipole number, $k$ the wavenumber of the gravitational wave and $L$ the distance to the pulsars. Then, we find that the overlap reduction function in the small angle limit approaches a value given by $\pi kL\,v^2\,(1-v^2)^2$ times a normalization factor, exactly matching the value for the autocorrelation recently derived. Although we focus on the $v<1$ case, our formulation is valid for any value of $v$.
\end{abstract}

\maketitle

\section{Introduction \label{sec:intro}}

The unique nature of pulsars, characterized by their remarkably stable rotational periods, allows for their utilization in a wide range of applications. These applications extend from the creation of Galactic maps to the exploration of Gravitational Waves (GWs) in the frequency range of $1-100\,\text{nHz}$. Of particular interest is the latter, an exciting prospect pioneered by Sazhin \cite{Sazhin:1978myk}, Detweiler \cite{Detweiler:1979wn}, Hellings and Downs~\cite{Hellings:1983fr}, and its subsequent analysis. Since then, substantial progress has been made, encompassing both phenomenological studies~\cite{Chen:2018znx,Reardon:2015kba,Taylor:2020zpk} and efforts to constrain theoretical models and test the predictions of General Relativity~\cite{Wu:2023rib,Finn:2001qi,Bernardo:2023mxc,deRham:2019ctd}.

With the rapid evolution of technology and expertise in recent years~\cite{vanHaasteren:2014qva,Lentati:2014hja,Ellis:2016mtg,Ellis:2013nrb}, the prospect of reliably measuring a stochastic GW background in the aforementioned frequencies has become increasingly feasible. This advancement is driven by techniques in the field of Pulsar Timing Array (PTA)~\cite{Taylor:2021yjx}. In fact, PTA collaborations recently reported $3-4\sigma$ evidence of a GW background \cite{NANOGrav:2023gor,NANOGrav:2023hvm,EPTA:2023fyk,Reardon:2023gzh,Xu:2023wog}. A key aspect of PTA measurements is the identification of the Overlap Reduction Function (ORF) \cite{Romano:2023zhb,AnilKumar:2023yfw,Liang:2023ary}, commonly denoted as $\Gamma_{ab}(\xi_{ab})$, where $\xi_{ab}$ is the angle between pulsars “$a$” and “$b$”. In General Relativity (GR), the ORF is known as the Hellings-Downs curve \cite{Hellings:1983fr}.

Interestingly, it has been shown that the ORF can also be used to test the phase velocity of the GWs \cite{Liang:2021bct,Liang:2023ary,Liang:2024mex,Cordes:2024oem}. A phase velocity of the GWs with $v\neq 1$ is a signature of modifications of gravity. For example, it is possible to have $v<1$ or $v>1$ in scalar-tensor theories of gravity \cite{Horndeski:1974wa,Deffayet:2009wt,Deffayet:2009mn,Deffayet:2011gz,Kobayashi:2011nu,Gleyzes:2014dya,Langlois:2015cwa,DeFelice:2018ewo,Gao:2014fra,Gao:2014soa} (see Refs.~\cite{Kobayashi:2019hrl,Lazanu:2024mzj} for recent reviews and Refs.~\cite{Gleyzes:2013ooa,Fujita:2015ymn,Crisostomi:2016tcp,BenAchour:2016cay,BenAchour:2016fzp,Takahashi:2017zgr,Babichev:2021bim} for an Effective Field Theory (EFT) perspective). In terms of metric transformations, $v\neq 1$ could also be motivated by disformal couplings to a scalar field \cite{Bekenstein:1992pj,Bettoni:2013diz,Zumalacarregui:2013pma}. In massive gravity one finds that the phase velocity is frequency dependent \cite{Finn:2001qi,LIGOScientific:2016lio,LIGOScientific:2016lio,LIGOScientific:2021sio,Liang:2021bct,Bernardo:2023mxc,Wang:2023div,Wu:2023rib,Bernardo:2023zna}. One can also use general EFTs to study modifications to the propagation speed and polarization of GWs \cite{deRham:2019ctd,deRham:2020zyh,Ezquiaga:2021ler,CarrilloGonzalez:2022fwg}. It is important to note that although the LIGO/VIRGO/KAGRA collaboration placed stringent constraints on the propagation speed of GWs \cite{LIGOScientific:2017zic,Ezquiaga:2017ekz} at frequencies of around $100\,{\rm Hz}$, it does not exclude departures from $v=1$ at lower frequencies, see, e.g., the arguments provided by Ref.~\cite{deRham:2018red}. Thus, PTAs provide a complementary probe to modified gravity. In general, one finds a different shape and frequency dependence of the ORF \cite{Liang:2021bct,Liang:2023ary,Liang:2024mex,Cordes:2024oem}, namely $\Gamma_{ab}(kL,\xi_{ab})$, where $k$ is the wavenumber of the GW and $L$ the distance to the pulsars. 

However, current analytical approximations to ORF in the $v<1$ case diverge at the small pulsar separation \cite{Liang:2023ary,Liang:2024mex,Cordes:2024oem}. Recently, Ref.~\cite{Cordes:2024oem} found an exact formula for the value of the autocorrelation, i.e., the correlation of timing residuals of a single pulsar, say $\Gamma_{aa}$, for general values of $v$. In particular, they provide the autocorrelation value for the $v<1$ case in the large $kL$ limit. However, there is the open question of how the ORF for $v<1$ approaches the autocorrelation value and why and when the analytical approximation breaks down.

\begin{figure}
\includegraphics[width=0.329\columnwidth]{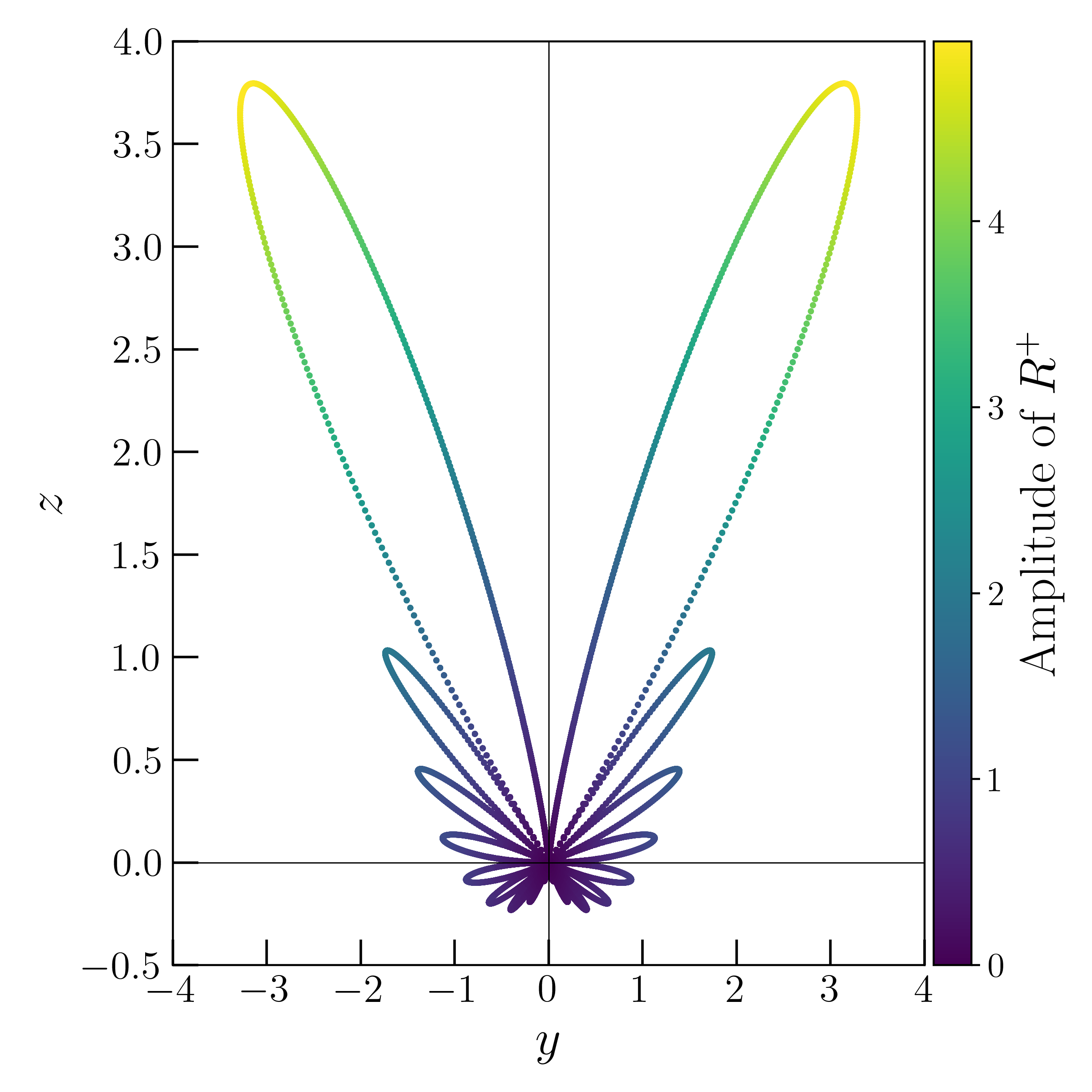}
\includegraphics[width=0.329\columnwidth]{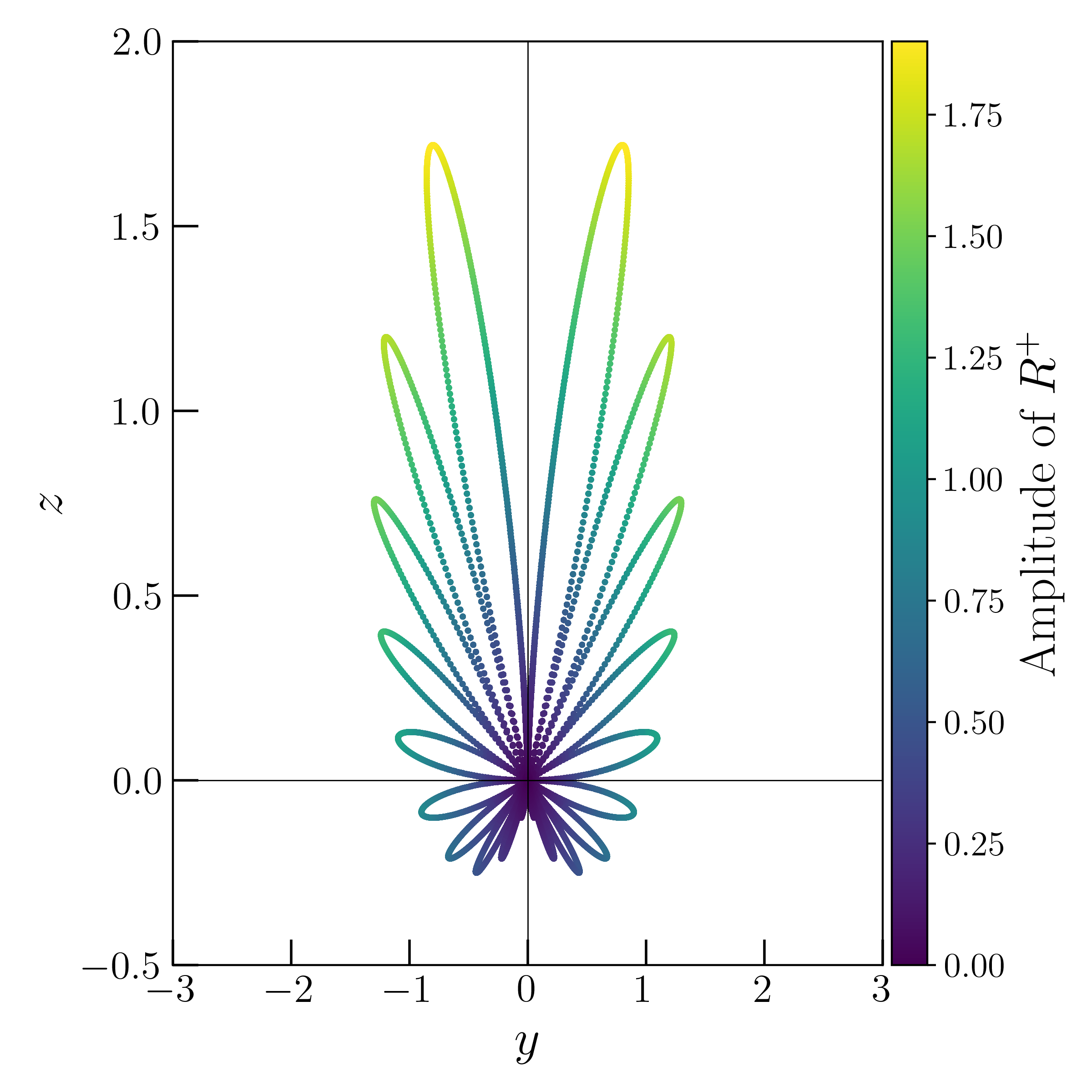}
\includegraphics[width=0.329\columnwidth]{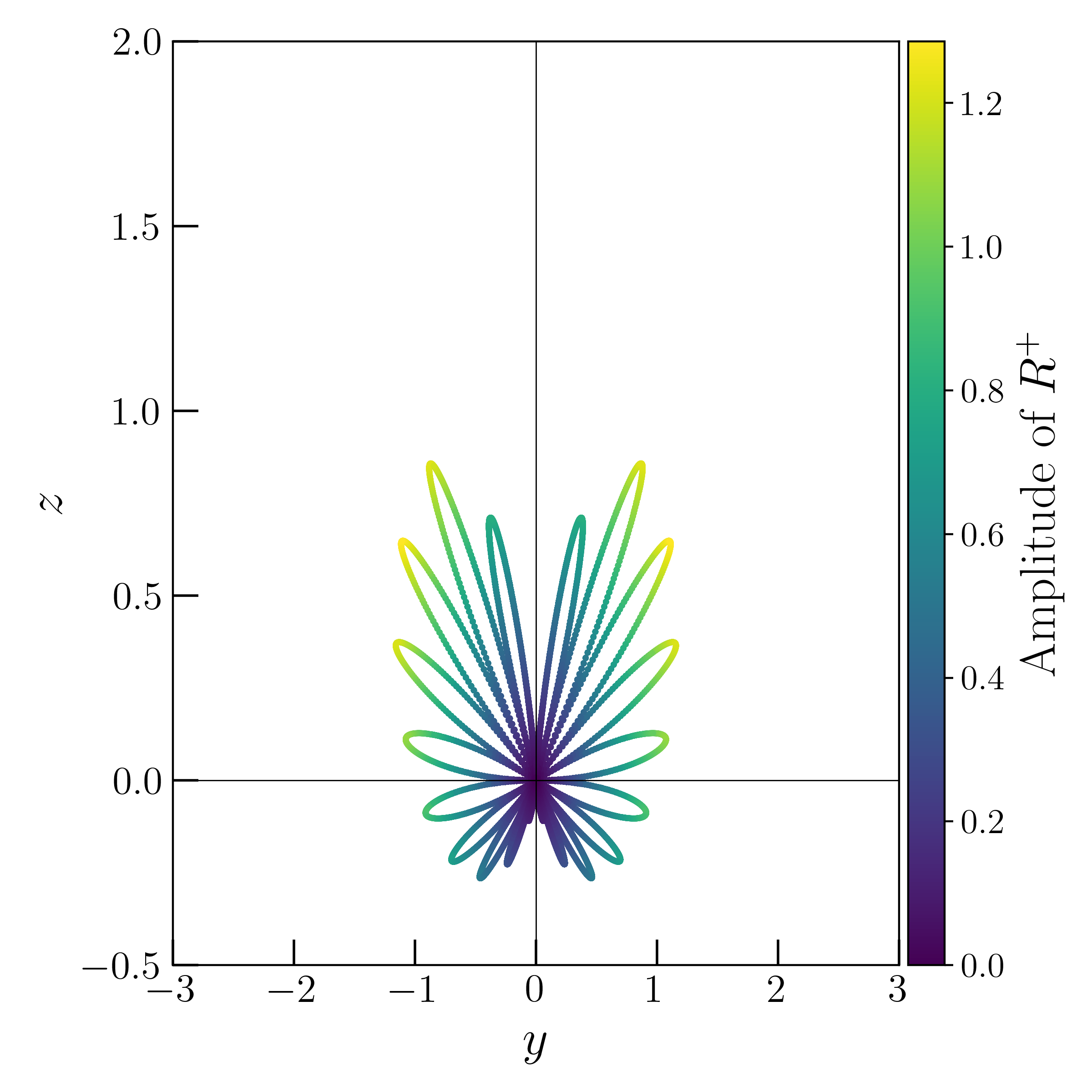}
\caption{Projection in the $y-z$ plane of the antenna response function \eqref{eq:antennaresponse} of PTAs for the plus polarization with the pulsar located in the $z$ direction for $v=\{0.8,1.0,1.2\}$ respectively from left to right. The color code shows the amplitude of the response function. For illustration purposes we fixed $kL=10$. Note that there is symmetry around the $z$ axis. See how for $v<1$ there is a particular direction of the source, i.e. when $\hat{\Omega}\cdot \hat{p}=-v$, where there is large redshift due to the “surfing” on the GWs. For $v=1$ the maximum response is when the direction source is close to the location pulsar but not right behind it. For $v>1$, the response already starts to decrease when the direction of source gets closer to the pulsar.}\label{fig:antenna} 
\end{figure}

The goal of this paper is to investigate the finite distance effects on the ORF with a special focus on the $v<1$ case. To do so, we work in the spherical harmonic decomposition and express the Legendre functions of the harmonic coefficients in terms of spherical Bessel functions. This connects us to the formulas that result from the Total Angular Momentum (TAM) formalism of Ref.~\cite{Dai:2012bc} (see also Refs.~\cite{Qin:2018yhy,Qin:2020hfy,Liu:2022skj,Bernardo:2022rif,Bernardo:2022xzl,Bernardo:2022vlj,Bernardo:2023pwt} for applications to PTAs, and Refs.~\cite{Qin:2020hfy,Bernardo:2022rif} in particular for the application to the $v<1$ case). As the spherical Bessel functions extend over the whole real line, they have, in our opinion, a more manageable dependence on the multipole number $\ell$. With our alternative formulation, we discover that the approximations derived in Refs.~\cite{Liang:2023ary,Liang:2024mex} are valid up to a cut-off at $\ell\sim \sqrt{1-v^2}\,kL$. Naively, a cut-off for $\ell\gtrsim kL$ is to be expected as the angular separation corresponding to the multipole number (roughly $\xi\sim 1/\ell$) is smaller than the typical features in the antenna response pattern (which is of the order of $1/kL$). Thus, multipoles higher than the cut-off do not contribute to the description of the antenna response. See Fig.~\ref{fig:antenna} for the antenna pattern for $v=\{0.8,1,1.2\}$.\footnote{We follow the procedure described in Ref.~\cite{Romano:2023zhb} with the antenna response modified to
\begin{align}\label{eq:antennaresponse}
R^A(k,\hat{\Omega})=\left(1-e^{-ikL(v+\hat{\Omega}\cdot\hat{p})}\right)\frac{v}{2}\frac{e^A_{ij}(\hat{\Omega})p_ip_j}{v+\hat{\Omega}\cdot\hat{p}}\,.
\end{align}
The first term inside the parenthesis is the so-called redshift transfer function and the rest is called geometrical projection factors. The subscript $A$ refers to plus $+$ or cross $\times$ polarization.
 We thank B.~Allen for the suggestion to draw the antenna pattern.} At the end of the paper, we show that such cut-off reconciles the ORF of Refs.~\cite{Liang:2021bct,Liang:2023ary} with the autocorrelation value of Ref.~\cite{Cordes:2024oem}.

This paper is organized as follows. In Sec.~\ref{sec:calculations}, we present an alternative formulation to the harmonic coefficients in terms of one integral of spherical Bessel functions. We dedicate Sec.~\ref{sec:approx} to study the behavior of the integral for $\ell<kL$ and $\ell> kL$. We show an exponential suppression in the latter case for $\ell> \sqrt{1-v^2}\,kL$. In Sec.~\ref{sec:ORF}, we compute the ORF for $v<1$ and show that it is finite in the small angle limit. We conclude our work in Sec.~\ref{sec:conclusion} with discussions and future directions. In App.~\ref{app:TAM} we show the equivalence of the equations in the TAM formalism and, in passing, provide a compact formula for their general integrals in the $kL\to\infty$ limit, Eq.~\eqref{eq:Inellinfinity}.

\section{Review and Re-formulation in terms of Bessel functions \label{sec:calculations}}

Our starting point is the expression of the Hellings-Downs curve in the spherical harmonic expansion, which following the notation of Ref.~\cite{Liang:2024mex} reads
\begin{align}\label{eq:HDharmonic}
\Gamma(kL,\xi)={\cal C}\sum_{\ell=2}^\infty a_\ell P_\ell(\cos\xi)\,,
\end{align}
where ${\cal C}$ is a normalization constant, which we will fix to ${\cal C}=3/64$ to compare with Ref.~\cite{Cordes:2024oem}, 
\begin{align}\label{eq:als}
a_\ell=(2\ell+1)\frac{2(\ell-2)!}{(\ell+2)!}|c_\ell|^2\,,
\end{align}
and
\begin{align}\label{eq:cloriginal}
c_\ell=\int_{-1}^1dx\frac{v}{v+x}\left(1-e^{-ikL(v+x)}\right)(1-x^2)^2\frac{d^2}{dx^2}P_\ell(x)\,.
\end{align}
We refer the reader to Refs.~\cite{Gair:2014rwa,Liang:2023ary,Romano:2023zhb,Allen:2024bnk} for more details on the derivation of the above formulas. In Eq.~\eqref{eq:cloriginal}, we denote with $v=\omega(k)/k$ the phase velocity of the GWs (where $\omega(k)$ is the dispersion relation), which in principle can have any positive value as well as $k$ dependence.\footnote{A $k$ dependence in the phase velocity appears in massive gravity \cite{Finn:2001qi,LIGOScientific:2016lio,LIGOScientific:2016lio,LIGOScientific:2021sio,Liang:2021bct,Bernardo:2023mxc,Wang:2023div,Wu:2023rib,Bernardo:2023zna} where $\omega^2=k^2+m_g^2$ with $m_g$ being the graviton’s mass.} It should be noted that we also assumed pulsars at equal distances for simplicity. The main goal of this section is to recast $c_\ell$ in a more suitable form for integration and analytical approximations. 

We start noting that the terms containing $v+x$ may be written in an integral form, namely
\begin{align}\label{eq:trick}
\frac{v}{v+x}\left(1-e^{-ikL(v+x)}\right)=i\int_0^{kL}dq \,e^{-i(v+x)q}\,.
\end{align}
We note that similar replacements are often referred to as “Feynman's trick”.
With the identity \eqref{eq:trick}, Eq.~\eqref{eq:cloriginal} becomes\footnote{We also used that
\begin{align}
\frac{d}{dx}P_\ell(x)=\frac{\ell+1}{1-x^2}\left(xP_\ell(x)-P_{\ell+1}(x)\right)\quad{\rm and}\quad (\ell+2)P_{\ell+1}(x)=(2\ell+3)xP_{\ell+1}(x)-(\ell+1)P_\ell(x)\,.
\end{align}
See Sec.~14.10 of Ref.~\cite{NIST:DLMF} for more details.
}
\begin{align}\label{eq:cl2}
c_\ell=iv(\ell+1)\int_0^{kL}dq \,e^{-ivq}\int_{-1}^1dx \,e^{-ixq}\left\{\left((\ell+2)x^2-\ell\right)P_\ell(x)-2xP_{\ell+1}(x)\right\}\,.
\end{align}
Now we can perform the $x$ integral by noting that (see Eq.18.17.19 in Ref.~\cite{NIST:DLMF})
\begin{align}
\int_{-1}^{1}dx e^{-ixq} P_\ell(x)=2 (-i)^\ell j_\ell(q)\,,
\end{align}
where $j_\ell(q)$ is the spherical Bessel of $\ell$-th order. Then, integrating Eq.~\eqref{eq:cl2} over $x$ and after some integration by parts,\footnote{We used that \begin{align}
\int_{-1}^1 dx \,x^n e^{-ixq} P_\ell(x)=2(-1)^\ell i^{\ell+n} j^{(n)}_\ell(t)\,,
\end{align}
where we indicate the $n$-th derivative with a superscript parenthesis.
} we obtain our main formula, which is given by
\begin{align}\label{eq:clnice}
c_\ell=2(-1)^\ell& \,i^{\ell+1}v(\ell+1)\Bigg\{\left((\ell+2)v^2-\ell\right){\cal I}_\ell(kL,v)-2iv\,{\cal I}_{\ell+1}(kL,v)\nonumber\\&
-e^{-ivkL}\left[\frac{\ell+ivkL}{kL}(\ell+2)j_\ell(kL)-\ell\,j_{\ell+1}(kL)\right]\Bigg\}\,,
\end{align}
where we used that $j_\ell(0)=0$ for $\ell>0$ and, for convenience, we introduced
\begin{align}\label{eq:mainintegral}
{\cal I}_\ell(kL,v)=\int_0^{kL}dq \,j_\ell(q)e^{-ivq}\,.
\end{align}
With the above procedure, we have reduced the calculation of $c_\ell$ given by Eq.~\eqref{eq:cloriginal} to a definite integral of a spherical Bessel function. This type of integrals appear in the TAM formalism (see App.~\ref{app:TAM}). Now, it all boils down to calculating ${\cal I}_\ell(kL,v)$, which unfortunately does not have an analytical solution in terms of known functions, as far as we are aware. It is interesting to note that the spherical Bessel functions in Eq.~\eqref{eq:clnice} can have an amplitude of ${\cal O}(\ell^2)$. Thus, for finite $kL$, cross terms in $|c_\ell|^2$ that include Bessel functions can introduce ${\cal O}(1)$ modulations on the amplitude of $a_\ell$. In the calculations of the ORF though, they mostly average out.

\section{Analytical approximations to the integral \texorpdfstring{${\cal I}_\ell(kL,v)$}{} \label{sec:approx}}

Before going into the detailed analytical approximations, it is instructive to look at Eq.~\eqref{eq:mainintegral} qualitatively. First, we note that in PTAs, we have GW frequencies between $f\sim 10^{-9}-10^{-8}\,{\rm Hz}$ and distances to pulsars of about $L\sim 1000\,{\rm lyr}$. It follows that typical values for $kL\sim 2\pi f L$ are of the order of $10^3-10^4$. We will then always consider that $kL\gg1$ in Eq.~\eqref{eq:mainintegral}. However, this is not enough to only consider the $kL\to\infty$ limit of the integral due to the behavior of the spherical Bessel function. This is because the spherical Bessel function, $j_\ell(q)$, first grows as $q^\ell$, peaks at around $q\sim \ell$, and continues to infinity with damped oscillations. This difference between $q>\ell$ and $q<\ell$ tells us that there will be a different behavior for $\ell\ll kL$ and $\ell\gg kL$.

To understand the change in behaviour of ${\cal I}_\ell(kL,v)$ \eqref{eq:mainintegral} in terms of $kL$, $\ell$ and $v$, it is instructive to take the large $\ell$ limit approximation of the spherical Bessel function given by
\begin{align}\label{eq:sphericalbesselrayleigh}
j_\ell(\mu\sec\beta)\approx\frac{1}{\mu}\sqrt{\frac{\cos\beta}{\tan\beta}}\cos\left(\mu(\beta-\tan\beta)+\tfrac{\pi}{4}\right)\,,
\end{align}
where $\mu=\ell+1/2$. The above equation is valid for $0<\beta<\pi/2$ and follows from the approximation to $J_\mu(x)$ derived by Rayleigh (see Sec.~8.22 of Ref.~\cite{Watson:1944:TBF}). Within such approximation we have that Eq.~\eqref{eq:mainintegral}, for $\ell<kL$, reads
\begin{align}\label{eq:approxintegral}
{\cal I}_\ell(\ell<kL,v)\approx\, &{\cal I}_\ell(\ell,v)\nonumber\\&+\int_0^{\beta_{kL}} \frac{d\beta}{2\mu}\sqrt{\frac{\tan\beta}{\cos\beta}}\left(e^{i\mu(\beta-\tan\beta)+i\tfrac{\pi}{4}}+e^{-i\mu(\beta-\tan\beta)-i\tfrac{\pi}{4}}\right)e^{-iv\mu\sec\beta}\,,
\end{align}
where $\beta_{kL}=\arccos(\ell/kL)$. In this form, it is clear that the integrand of Eq.~\eqref{eq:approxintegral} contains two highly oscillating functions with frequency proportional to $\mu$ (or $\ell$). 

We can gain insight into the behavior of Eq.~\eqref{eq:approxintegral} by looking for saddle points of the imaginary part of the integrand. For $v<1$, we find that the saddle point of the product of the oscillations lies at $\sin\beta=v$. This means that the integrand contributes most around $q\sim\mu/\sqrt{1-v^2}$, which for small values $v$ is, unfortunately, in the regime where Eq.~\eqref{eq:sphericalbesselrayleigh} is not very reliable, at least regarding the amplitude. Nevertheless, approximating the integral by the integrand at that point captures the essence of the integral. Namely, it will be an oscillating function of $\ell$ for a given $v<1$ (actually, the integrand resembles a Hankle function of zeroth order as it goes as $e^{-i\mu(\arcsin v+\pi/4)}$; something close to what we shall find later). In the opposite regime, namely $v>1$, the saddle point is complex. If we say that $\beta=a+ib$, the saddle point is located at $a=\pi/2$ and $\cosh b=v$ (at the other extreme of the real $\beta$ line). This means that the integral is exponentially suppressed with $\ell$, roughly as $e^{-\mu\arccosh v}$. Both behaviors, for $v<1$ and $v>1$ are consistent with the $kL\to\infty$ solutions to ${\cal I}_\ell$ \eqref{eq:mainintegral}, given by Eq.~\eqref{eq:Iinfinity}. We thus conclude that the saddle point argument, though not completely rigorous, implies that for $v<1$, no “constructive interference” occurs when
\begin{align}\label{eq:lcut}
\ell\gtrsim\ell_{\rm cut}=\sqrt{1-v^2} kL\,.
\end{align}
Note that, although in Ref.~\cite{Liang:2024mex} it was argued that $c_\ell$ for $\ell\gg kL$ should be exponentially suppressed, we derived the position of the cut-off and will derive the precise exponential suppression. For $v>1$ the behavior change might happen at lower $\ell$ as the saddle point corresponded to larger values of $q$. However, this will not be relevant for the discussion of the ORF as, in this case, only the lower multipoles contribute. We also note that for $q<\ell$, the spherical Bessel function oscillates no more, and both cases should have a similar behavior. We proceed to study the integral in two different limits and derive the precise scaling of the integral \eqref{eq:mainintegral} with $\ell$ and $kL$.

\subsection{Analytical approximation in the \texorpdfstring{$\ell< kL$}{} regime \label{sec:approxlargekL}}

We start by looking at the $\ell\ll kL$ limit to Eq.~\eqref{eq:mainintegral}, which corresponds to the case when there are many GW oscillations within the angular scale corresponding to $\ell$. So, effectively we may send $kL\to\infty$. In this limit, there are closed formulas in terms of Legendre polynomials.\footnote{These formulas can be derived using the formulas in Secs.~10.22 and 14.3 of Ref.~\cite{NIST:DLMF}, in particular using Eq. 10.22.56 and the hypergeometric function representation of the Legendre polynomials. Alternatively, one can ask a software like \textsc{Mathematica}.} Explicitly, we find that
\begin{align}\label{eq:Iinfinity}
{\cal I}_\ell(kL\to\infty,v)=\left\{
\begin{aligned}
&\frac{\pi}{2}(-i)^\ell P_{\ell}(v)+(-i)^{\ell+1}Q_{\ell}(v)&(v<1)\\
&(-i)^{\ell+1}{\cal Q}_\ell(v)& (v>1)
\end{aligned}
\right.\,,
\end{align}
where $Q_{\ell}(v)$ is the Legendre polynomial of the second kind for $v<1$, also known as Ferrers function, and ${\cal Q}_\ell(v)$ is the Legendre function for $v>1$. It is interesting to note that, in our formulation, the difference between $v>1$ and $v<1$ comes from whether the frequency of the oscillations of the exponential is fast enough to interfere with the oscillations of the Bessel function (see also the discussion on the saddle point approximation at the beginning of the section). In the original formulation, the difference between the $v>1$ and $v<1$ could be associated with the seeming pole at $x=-v$ \cite{Liang:2023ary}. Regarding the Hypergeometric representation of Legendre functions, the values $v<1$ and $v>1$ correspond to a different choice of branch cuts. Also note that, for the same reason, the integral Eq.~\eqref{eq:mainintegral} is not well-defined for $v=1$ in the $kL\to\infty$ limit, though $c_\ell$ in \eqref{eq:clnice} is. Now, with Eq.~\eqref{eq:Iinfinity}, we find that Eq.~\eqref{eq:clnice}, in the $\ell\ll kL$ regime, is approximated by
\begin{align}\label{eq:clnice2}
c_\ell&(\ell\ll kL)\approx\,4(-1)^\ell\nonumber\\&\times \left\{
\begin{aligned}
&\frac{v(\ell+1)}{2}\left[\left((\ell+2)v^2-\ell\right)\left(Q_\ell(v)+\tfrac{i\pi}{2}P_\ell(v)\right)-2v\left(Q_{\ell+1}(v)+\tfrac{i\pi}{2}P_{\ell+1}(v)\right)\right]& (v<1)\\
&1& (v=1)\\
&\frac{v(\ell+1)}{2}\left[\left((\ell+2)v^2-\ell\right){\cal Q}_\ell(v)-2v{\cal Q}_{\ell+1}(v)\right]& (v>1)
\end{aligned}
\right.\,,
\end{align}
which coincides with the results of Ref.~\cite{Liang:2024mex}.\footnote{After using that $P_\ell(-x)=(-1)^{\ell}P_\ell(x)$ and $Q_\ell(-x)=(-1)^{\ell+1}Q_\ell(x)$.} For the limit of $v\to1$ we took the total expression in terms of Legendre polynomials and used the asymptotic formulas near the singularity.\footnote{For example, we have that $P_\ell(x\to 1^{-})\to 1$ and $Q_\ell(x\to 1^{-})\to \frac{1}{2}\ln\left(\frac{2}{1-x}\right)-\gamma_E-\psi(\ell+1)$, where $\psi(\ell)$ is the digamma function and $\gamma_E\approx 0.577$ Euler’s constant.} All divergences cancel, and we obtain the well-known result for $v=1$. As a curiosity we note that the combinations of Legendre polynomials in Eq.~\eqref{eq:clnice2}, can be written in terms of single associated Legendre polynomials, concretely in terms of $P^{-2}_\ell(v)$ and $Q^{-2}_\ell(v)$ (see App.~\ref{app:TAM}, Eq.~\eqref{eq:Inellinfinity}). The same applies to the $v>1$ case, namely the result is proportional to ${\cal Q}^{-2}_{\ell}(v)$. In terms of the associated Legendre functions, the limit $v\to 1$ is smooth.

\subsection{Analytical approximation in the \texorpdfstring{$\ell\gtrsim kL$}{} regime \label{sec:approxlargel}}

We now turn to the $\ell>kL$ regime, which corresponds to the case when the angular resolution associated with the multipole number is finer than the number of GW oscillations. This is the regime where we expect finite distance effects to take place. Noting that the $\ell>kL$ regime roughly corresponds to the regime where the spherical Bessel function does not oscillate, we use the power series expression for $j_\ell(q)$, namely
\begin{align}\label{eq:powerseries}
j_{\ell}\left(q\right)=\frac{\sqrt{\pi}}{2}\sum_{m=0}^{\infty}\frac{(-1)^{m}}{m!\,\Gamma\left[\ell+m+3/2\right]}\left(
\frac{q}{2}\right)^{\ell+2m}\,.
\end{align}
Using the expansion Eq.~\eqref{eq:powerseries}, we can formally write Eq.~\eqref{eq:mainintegral} as 
\begin{align}\label{eq:expansionofintegral}
{\cal I}_{\ell}(kL,v)={\sqrt{\pi}}&\sum_{m=0}^{\infty}\frac{(-1)^{m}\left(
2v\right)^{-1-\ell-2m}}{m!\,\Gamma\left[\ell+m+3/2\right]}\left({\rm Ci}_{\ell+2m+1}\left[vkL\right]-i\,{\rm Si}_{\ell+2m+1}\left[vkL\right]\right)\,,
\end{align}
where ${\rm Ci}_{n}[z]=\int_{0}^{z}t^{n-1}\cos t\,dt$ and ${\rm Si}_{n}[z]=\int_{0}^{z}t^{n-1}\sin t\,dt$ respectively are the generalized sine and cosine integrals.\footnote{See Sec.~8.21 of Ref.~\cite{NIST:DLMF} for more details.} In the current form, there is no resummation formula for Eq.~\eqref{eq:expansionofintegral}, to the best of our knowledge. However, we can resum Eq.~\eqref{eq:expansionofintegral} under a series of approximations.

We first note that, both ${\rm Ci}_{n}[z]$ and ${\rm Si}_{n}[z]$, where in our set up $z=vkL$, decrease similarly with decreasing $z$ for $z\ll n$, namely ${\rm Ci}_{n}[z]\propto z^n/n$ and ${\rm Si}_{n}[z]\propto z^n/(1+n)$.\footnote{For $z\ll 1$ we have that ${\rm Si}_{n}[z]\propto z^{n+1}$. However, this corresponds to the case where $vkL\ll1$, where a good approximation is to neglect the exponential in Eq.~\eqref{eq:mainintegral} as it barely oscillates. Then Eq.~\eqref{eq:mainintegral} is well approximated by \begin{align}
    \int_0^{kL}dq j_\ell(q)=\frac{\sqrt{\pi }}{(l+1) \Gamma \left[\ell+3/2\right]}\left(\frac{kL}{2}\right)^{l+1}\,
   _1F_2\left(\tfrac{l+1}{2};\tfrac{l+3}{2},l+\tfrac{3}{2};-\left(\tfrac{kL}{2}\right)^2\right)\,.
\end{align}} we also notice that the remaining factor is an oscillatory function slowly decaying with $z$. For the ${\rm Ci}_{n}[z]$ it is initially very similar to $\cos(z)$ and for the ${\rm Si}_{n}[z]$ to $\sin(z)$. We then use the approximation that
\begin{align}
{\rm Ci}_{n}[z\ll n]\approx \frac{z^n}{n}\cos(z)\quad{\rm and}\quad {\rm Si}_{n}[z\ll n]\approx \frac{z^n}{1+n}\sin(z)\,.
\end{align}
Then, we find that
\begin{align}\label{eq:hypergeometricsum}
{\cal I}_{\ell}(kL,v)&\approx{\sqrt{\pi}}\sum_{m=0}^{\infty}\frac{(-1)^{m}\left(
2kL\right)^{1-\ell-2m}}{m!\,\Gamma\left[\ell+m+3/2\right]}\left(\frac{\cos(vkL)}{\ell+2m+1}-\frac{i\sin(vkL)}{\ell+2m+2}\right)\nonumber\\&
=\frac{\sqrt{\pi }}{(l+1) \Gamma \left[\ell+3/2\right]}\left(\frac{kL}{2}\right)^{l+1}\Bigg\{\,
   _1F_2\left(\tfrac{l+1}{2};\tfrac{l+3}{2},l+\tfrac{3}{2};-\left(\tfrac{kL}{2}\right)^2\right)\cos(vkL)\nonumber\\&\qquad\qquad\qquad\qquad-\frac{i\sin(vkL) (l+1)}{l+2} \,
   _1F_2\left(\tfrac{l}{2}+1;\tfrac{l}{2}+2,l+\tfrac{3}{2};-\left(\tfrac{kL}{2}\right)^2\right)\Bigg\}\,,
\end{align}
    where $_1F_2\left(a;b,c;-x^2\right)$ is the generalized Hypergeometric function. We note that this approximation is close to considering a constant phase in the integral Eq.~\eqref{eq:mainintegral} and integrating the spherical Bessel function only. In fact, since $v<1$, we expect that the approximation should still be valid for $\ell>vkL$, which goes below the cut-off at $\ell \sim kL$. Thus, this approximation is particularly good for the case with $v\ll1$.

In the current form, Eq.~\eqref{eq:hypergeometricsum} is not very enlightening. So, let us discuss some interesting limits. First, keeping the leading order term in $kL/\ell$ in a Taylor series expansion yields
\begin{align}\label{eq:Ileadingorder}
{\cal I}_{\ell}(kL,v)&\approx\frac{\sqrt{\pi }}{(\ell+1) \Gamma \left[\ell+3/2\right]}\left(\frac{kL}{2}\right)^{\ell+1}\left(\cos(vkL)-\frac{i (l+1)}{l+2}\sin(vkL)\right)\nonumber\\&\approx \frac{e^\ell}{\sqrt{2} \ell}\left(\frac{kL}{2\ell}\right)^{\ell+1}e^{-ivkL}\,.
\end{align}
Note that, despite the exponential dependence on $\ell$, Eq.~\eqref{eq:limitbessels} quickly decays for $\ell\gg kL$.
In the transitional regime, where $\ell \gtrsim kL$, the $kL/\ell$ dependence can be inferred from the following. First, we note that for $\ell \gtrsim kL$, the integrand in Eq.~\eqref{eq:mainintegral} peaks at the upper integration limit, that is, at $q=kL$. We then approximate the integral by the value of the integrand at the maximum, which gives ${\cal I}_\ell(kL\lesssim \ell,v)\sim \,j_\ell(kL)e^{-ivkL}$ up to ${\cal O}(1)$ errors. We then use the approximation formula in transitional regions to $j_\ell(kL)$ given by Nicholson (see Sec.~8.43 of Ref.~\cite{Watson:1944:TBF}), which reads
\begin{align}
j_\ell(\ell\gtrsim kL)&\approx \sqrt{\frac{2}{3\pi}}\frac{\sqrt{\nu-kL}}{kL}K_{\frac{1}{3}}\left[\frac{2 \sqrt{2}
   (\nu-kL)^{3/2}}{3 \sqrt{kL}}\right]\nonumber\\&\approx \frac{1}{2^{3/4}\ell(1-kL/\ell)^{1/4}}\left(\frac{\ell}{kL}\right)^{3/4}e^{-\ell\sqrt{\frac{\ell}{kL}}\frac{2 \sqrt{2}(1-kL/\ell)^{3/2}}{3}}\,,
\end{align}
where $\nu=\ell+1/2$, $K_{1/3}[x]$ is the modified Bessel function of the second kind and in the last step we took the large $\ell$ limit and we approximated $K_{1/3}[x]\sim \sqrt{\tfrac{\pi}{2x}}e^{-x}$. Thus, for $\ell \gtrsim kL$, the initial decay of Eq.~\eqref{eq:mainintegral} is proportional to $e^{-\ell\sqrt{{\ell}/{kL}}}$, which is close to the behaviour speculated in Ref.~\cite{Liang:2024mex}. For $\ell \gg kL$ then it falls off as $e^\ell\left({kL}/{\ell}\right)^{\ell+1}$. In fact,  for practical purposes, we find that the approximation given by
\begin{align}\label{eq:Ihankel}
{\cal I}_{\ell}(\ell\gtrsim kL,v)&\approx \frac{e^{-ivkL}}{2^{3/4}\ell(1-kL/\ell)^{1/4}}\left(\frac{\ell}{kL}\right)^{3/4}e^{-\ell\sqrt{\frac{\ell}{kL}}\frac{2 \sqrt{2}(1-kL/\ell)^{3/2}}{3}}\,,
\end{align}
already gives a good approximation to the integral in the relevant range of $\ell \gtrsim kL$, and it is much less numerically expensive than Eq.~\eqref{eq:hypergeometricsum}. Though the approximation is not good for $\ell \gg kL$, the integral is already several orders of magnitude smaller and does not contribute to the Hellings-Downs curve \eqref{eq:HDharmonic}.

\begin{figure}
\includegraphics[width=0.49\columnwidth]{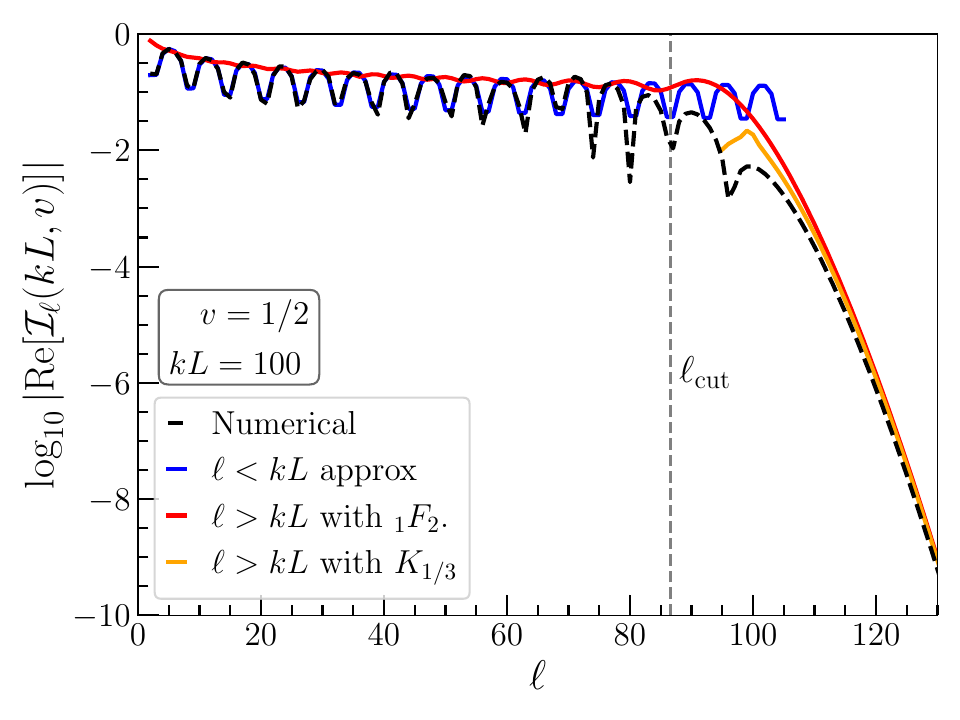}
\includegraphics[width=0.49\columnwidth]{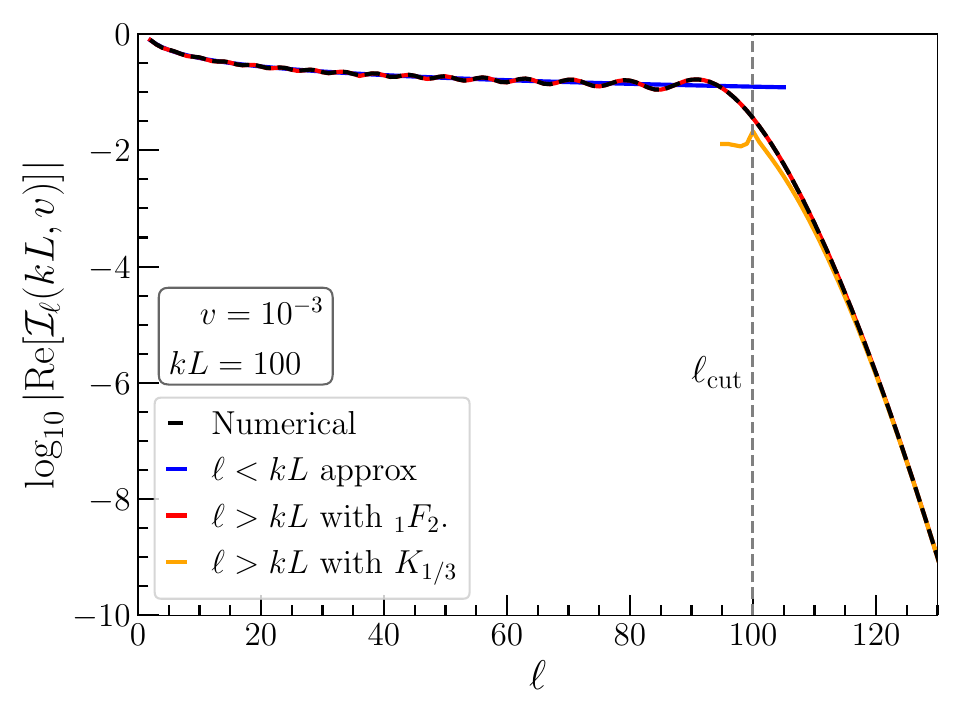}
\includegraphics[width=0.49\columnwidth]{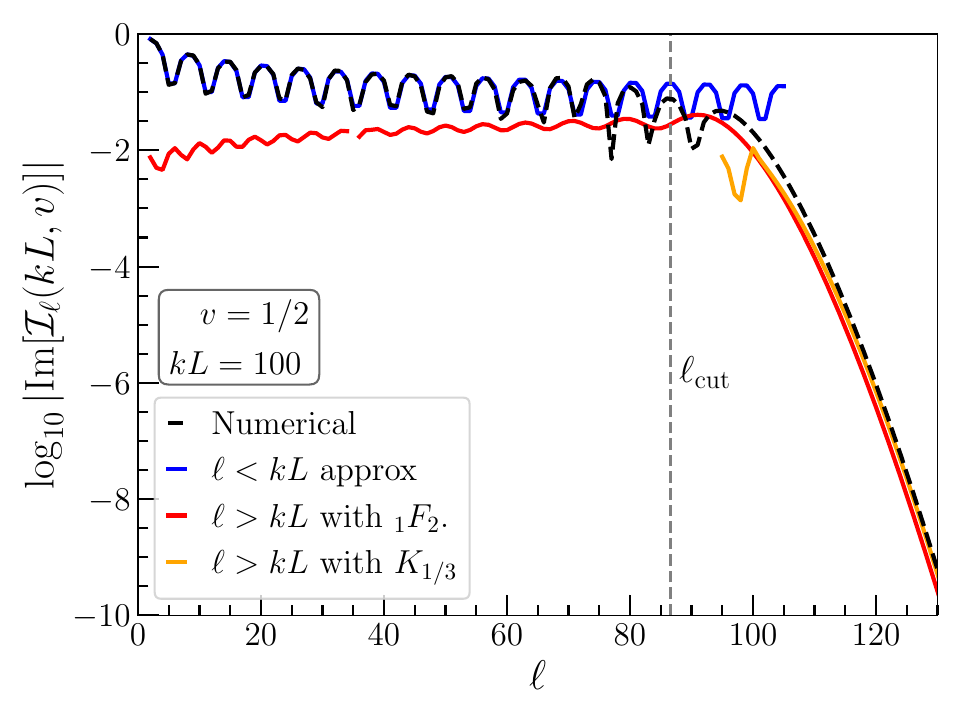}
\includegraphics[width=0.49\columnwidth]{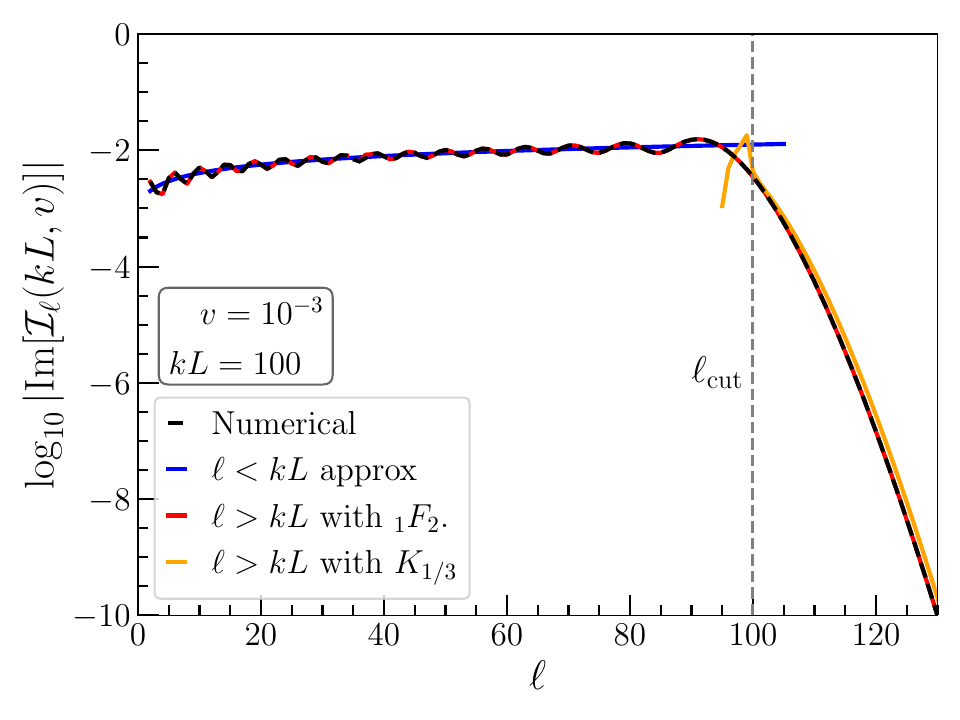}
\caption{Integral ${\cal I}_\ell(kL,v)$ in logarithmic scale for a fixed $kL=100$ as function of $\ell$. The top and bottom panels, respectively, show the real and imaginary parts of the integral. Vertical gray dashed lines show the location of $\ell_{\rm cut}$ \eqref{eq:lcut}. On the left and right panels, we show the cases of $v=1/2$ and $v=10^{-3}$ as two representatives of extreme cases. In black dashed lines, we show the numerical results. In blue, we show the approximation for $\ell<kL$ given by Eq.~\eqref{eq:Iinfinity}. In orange and red, we show the approximations for $\ell>kL$. The red line shows the approximation in terms of Hypergeometric functions \eqref{eq:hypergeometricsum}, while the orange one shows the qualitative approximation in terms of the modified Bessel function \eqref{eq:Ihankel}. We extrapolate the red line to the $\ell<kL$ regime to show that for smaller values of $v$, the approximation \eqref{eq:hypergeometricsum} is also good there. Note how the qualitative approximation by the modified Bessel function \eqref{eq:Ihankel} describes well the exponential decay. Also, note how the exponential suppression starts at around $\ell_{\rm cut}\sim \sqrt{1-v^2}kL$. The effect of $\sqrt{1-v^2}$ is slightly more apparent in the case of $v=1/2$.}\label{fig:Il} 
\end{figure}

We show in Fig.~\ref{fig:Il} the values of the integral ${\cal I}_\ell(kL,v)$ for $v=1/2$ and $v=10^{-3}$ with $kL=100$. We choose a relatively low value of $kL$ for illustration purposes. Later, we will fix $kL=1000$ to estimate the ORF numerically. In Fig.~\ref{fig:Il} we show the numerical results together with the approximation \eqref{eq:Iinfinity} for $\ell<kL$ and both Eqs.~\eqref{eq:hypergeometricsum} and \eqref{eq:Ihankel} for $\ell>kL$. See how the qualitative approximation given by the modified Bessel function, Eq.~\eqref{eq:Ihankel}, describes well the exponential decay of the integral for $\ell>kL$. We also checked numerically that the Taylor series expansion of Eq.~\eqref{eq:hypergeometricsum}, namely Eq.~\eqref{eq:Ileadingorder}, describes well the decay but only for $\ell\gg kL$. It gives a bad approximation for $\ell$ near $kL$. We conclude that the integral ${\cal I}_\ell(kL,v)$ is well approximated by the large $kL$ approximation \eqref{eq:Iinfinity} and a cut-off at $\ell\sim \sqrt{1-v^2}kL$ (see the discussion around Eq.~\eqref{eq:lcut}). We also checked numerically that the value of $\ell_{\rm cut}$ describes well the start of the decay.

We end this section by noting that the calculations in this subsection are valid for any value of $v$. Nevertheless, for $v\geq 1$, the summation in the Hellings-Downs curve, Eq.~\eqref{eq:HDharmonic}, converges quickly enough so that the effects are negligible.

\section{Finite distance effects to the ORF \label{sec:ORF}}

Here we compute the ORF \eqref{eq:HDharmonic} and focus on the coincident pulsar limit, where previous approximations gave a divergent result. We start by computing $a_\ell$ given by \eqref{eq:als}, which contains $|c_\ell|^2$. For numerical calculations, we use the full $c_\ell$ given by \eqref{eq:clnice}. For the analytical approximations we neglect the Bessel functions in \eqref{eq:clnice} and take
\begin{align}\label{eq:clnice3}
|c_\ell|^2\approx 4v^2(\ell+1)^2\,\Big|\left((\ell+2)v^2-\ell\right){\cal I}_\ell(kL,v)-2iv\,{\cal I}_{\ell+1}(kL,v)\Big|^2\,.
\end{align} 
Though we neglect the Bessel functions, it is interesting to note that for $\ell\gg kL$, a Taylor series expansion of the Bessel term in \eqref{eq:clnice} yields
\begin{align}\label{eq:limitbessels}
\frac{\ell+ivkL}{kL}(\ell+2)j_\ell(kL)-\ell\,j_{\ell+1}(kL)\approx \frac{\sqrt{\pi } (l+2) (kL)^{\ell-1} \ell}{ 2^{\ell+1}\Gamma
   \left(l+\frac{3}{2}\right)}\approx \frac{e^\ell}{2^{\ell+\frac{3}{2}}} \left(\frac{kL}{\ell}\right)^{\ell-1}\,,
\end{align}
where in the last step, we took the large $\ell$ limit. It is interesting to note that in this limit, the contribution from the spherical Bessel functions, Eq.~\eqref{eq:limitbessels} decays with two powers of $(kL/\ell)$ less than Eq.~\eqref{eq:Ileadingorder}. We confirm numerically that the Bessel contribution to $c_\ell$ quickly dominates for $\ell> kL$. This is nevertheless unimportant as the suppression is huge, and it does not contribute to the ORF.

Because in Sec.~\eqref{sec:calculations} we showed that $c_\ell$ is exponentially suppressed for $\ell>\sqrt{1-v^2}\,kL$, we consider for analytical purposes that the ORF \eqref{eq:HDharmonic2} for $v<1$ is well approximated by
\begin{align}\label{eq:HDharmonic2}
\Gamma(kL,\xi)={\cal C}\sum_{\ell=2}^{\ell_{\rm cut}} a_\ell P_\ell(\cos\xi)\,,
\end{align} 
where we introduced a cut-off to the sum at $\ell_{\rm cut}=\sqrt{1-v^2}\,kL$.\footnote{Strictly speaking the sum is from $\ell=2$ until the integer number closer to $\ell_{\rm cut}$. For large $\ell$ this does not make a difference.} Thus, we only need to know the value of $a_\ell$ in the $\ell\ll kL$ regime. As shown in Ref.~\cite{Liang:2024mex}, the large $\ell$ limit of $c_{\ell}$ in the case of $v<1$ yields\footnote{We used that $P_\ell(\cos\theta)\approx \sqrt{\theta/\sin\theta}J_0\left(\left(\ell+\tfrac{1}{2}\right)\theta\right)$ and $Q_\ell(\cos\theta)\approx -\tfrac{\pi}{2}\sqrt{\theta/\sin\theta}Y_0\left(\left(\ell+\tfrac{1}{2}\right)\theta\right)$. }
\begin{align}\label{eq:clnice4}
c_\ell(\ell\ll kL;\ell\gg1)\approx\,(-1)^\ell i\pi&
v(\ell+1)\sqrt{\frac{\theta}{\sin\theta}}\nonumber\\&\times\left[\left((\ell+2)v^2-\ell\right)H_0^{(1)}\left(\left(\ell+\tfrac{1}{2}\right)\theta\right)-2vH_0^{(1)}\left(\left(\ell+\tfrac{3}{2}\right)\theta\right)\right]\,,
\end{align}
where $\cos\theta=v$ and $H_0^{(1)}(x)$ is the Hankle function of the first kind. Further using that in the $\ell\gg1$ limit $H_0^{(1)}(x\gg1)\approx\sqrt{\tfrac{2}{\pi x}}e^{i(x-\pi/4)}$ and that the prefactor in $a_\ell$ \eqref{eq:als} goes as $4/\ell^3$, one obtains that
\begin{align}\label{eq:alnice}
a_\ell(\ell\ll kL;\ell\gg1)\approx 8\pi v^2(1-v^2)^{3/2}\,.
\end{align}
Namely, for large enough $\ell$ the coefficients $a_\ell$ are constant. 

\begin{figure}
\includegraphics[width=0.49\columnwidth]{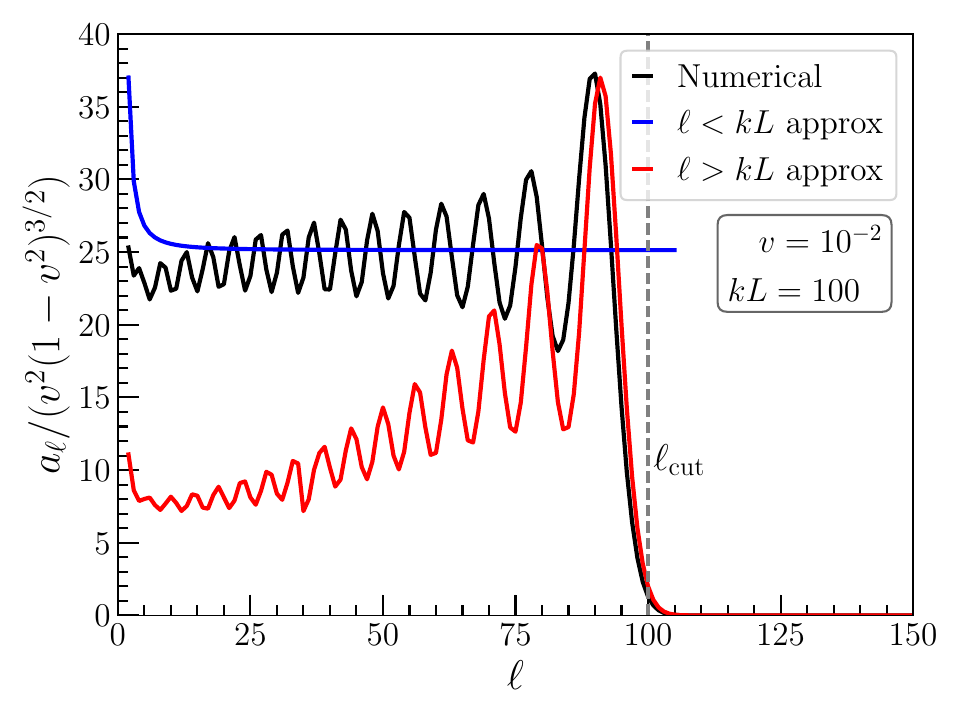}
\includegraphics[width=0.49\columnwidth]{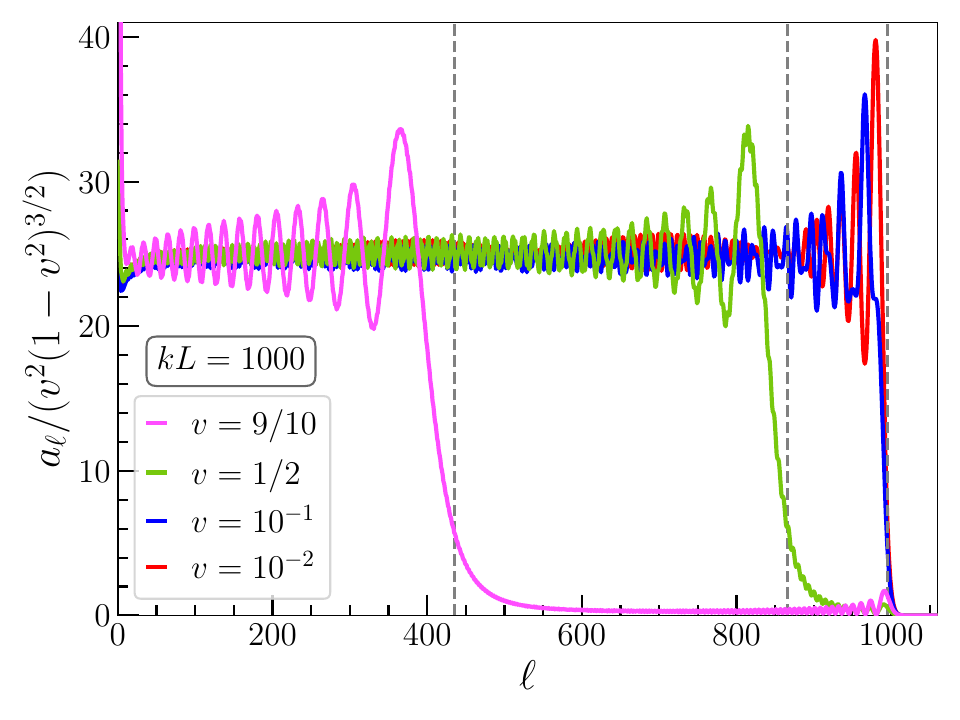}
\caption{Harmonic coefficients $a_\ell$ in Eq.~\eqref{eq:HDharmonic} as a function of $\ell$ for a fixed value of $kL$ and normalized by $v^{-2}(1-v^2)^{-3/2}$ so as to have the same amplitude independent of $v$. Vertical gray dashed lines indicate the location of $\ell_{\rm cut}$ \eqref{eq:lcut}. On the left panel, we show the case of $kL=100$ and $v=10^{-2}$ for illustration purposes. In black, we show the numerical result, and in blue and red, we show the $\ell<kL$ and $\ell>kL$ approximations given by only using Eqs.~\eqref{eq:Iinfinity} and \eqref{eq:hypergeometricsum}, respectively. For smaller values of $v$, the approximation for $\ell>kL$ becomes better. The oscillations in the black lines for $\ell<kL$ come from the Bessel functions in \eqref{eq:clnice}, which we omitted in the analytical approximation. They average out, though, in the ORF. On the right panel, we show the numerical results for $kL=1000$ for $v=\{9/10,1/2,10^{-1},10^{-2}\}$ respectively in magenta, green, blue, and red lines. The vertical dashed line corresponds from left to right to $\ell_{\rm cut}$ for $v=\{9/10,1/2\}$ and the $v\ll1$ limit, i.e. $\ell_{\rm cut}\sim kL$. See how the cut-off to $a_\ell$ is roughly at $\ell_{\rm cut}\sim \sqrt{1-v^2}kL$. }\label{fig:al} 
\end{figure}

We show in Fig.~\ref{fig:al} the $a_\ell$ for the case of $kL=100$ and $kL=1000$, respectively, on the left and right panels. The former is for illustration purposes; the latter is of the order of current observations.  In the left panel, we also fixed $v=10^{-2}$ as a small $v$ example, but the approximations work well in other cases, too. The exception is $v\gtrsim 1/3$ where the cut-off in $\ell\sim \sqrt{1-v^2}kL$ is noticeably different than $kL$. In that case, the $\ell>kL$ approximation \eqref{eq:hypergeometricsum} gives the right exponential decay but shifted to $\ell=kL$. For smaller values of $v$, the approximations get better, and Eq.~\eqref{eq:hypergeometricsum} is relatively good for some $\ell$ inside the $\ell<kL$ regime. In the right panel, we show the case of $kL=1000$ for $v=\{9/10,1/2,10^{-1},10^{-2}\}$. There it is clear that there is a very sharp cut-off roughly at $\ell\sim \sqrt{1-v^2}\,kL$. Let us remind the reader that value for $\ell_{\rm cut}$ is analytically derived in the beginning of Sec.~\ref{sec:calculations}. Though one could try to numerically find an effective cut-off, we will see shortly that $\ell_{\rm cut}$ given by Eq.~\eqref{eq:lcut} recovers the value of the autocorrelation function.

Now, let us discuss the impact of Eq.~\eqref{eq:alnice} on the ORF. In Ref.~\cite{Liang:2023ary}, the constant approximation is taken up to $\ell\to\infty$, which, using the resummation formulas of the Legendre polynomials, yields a factor $1/\sqrt{1-\cos\xi}$. However, this is divergent in the $\xi\to 0$ limit, as noted in Refs.~\cite{Liang:2023ary,Liang:2024mex,Cordes:2024oem}. Physically speaking, divergence is due to the fact in the strict $kL\to\infty$ limit, the pulsars are infinitely far away, and the “surfing” \cite{Liang:2024mex} on top of the GW takes place infinitely. In our terminology, the cut-off is sent to $\ell_{\rm cut}\to \infty$, and the sum diverges proportionally to $\ell_{\rm cut}$. However, for finite values of $kL$, we have that the value of ORF \eqref{eq:HDharmonic2} in the $\xi\to0$ limit is finite and given by 
\begin{align}\label{eq:HDharmonic3}
\Gamma(kL,\xi=0)=\frac{3}{64}\sum_{\ell=2}^{\ell_{\rm cut}} a_\ell\,,
\end{align}
where we set the normalization constant to $3/64$ for comparison with Ref.~\cite{Cordes:2024oem}.
We give an analytical estimate as follows. Using that $kL\gg1$ and so that $\ell_{\rm cut}\gg1$, we note that most of the contribution to the coincident limit comes from the large $\ell$ sum. We then assume that $a_{\ell}$ is constant for all $\ell$ and given by Eq.~\eqref{eq:alnice}, which yields that
\begin{align}\label{eq:gammaestimate}
\Gamma(kL,\xi\to0)\approx\frac{3}{64}a_\ell \ell_{\rm cut}=\frac{3\pi}{8} \,kL \,v^2(1-v^2)^2\,.
\end{align}
We numerically checked for $kL=1000$ that this limit is accurate within $0.1\%$. This further supports our analytical estimation.\footnote{It is interesting that $\Gamma(kL,\xi\to0)$ is not necessarily large despite the fast growth for small but not too small $\xi$. In fact, for $v<1/\sqrt{kL}$, it achieves values less than in the GR case. This limit is, unfortunately, not so interesting from the observational point of view, as in PTAs the statistical significance for $v<0.4$ drops below $3\sigma$ as shown in Ref.~\cite{Liang:2024mex} using the CPTA data.} We also note that our limit for $\xi\to0$ of the ORF matches exactly the value of the auto-correlation obtained in Ref.~\cite{Cordes:2024oem}; see their Eq.~(28)  with the identification of $vkL=2\pi fL$. 

\begin{figure}
\includegraphics[width=0.49\columnwidth]{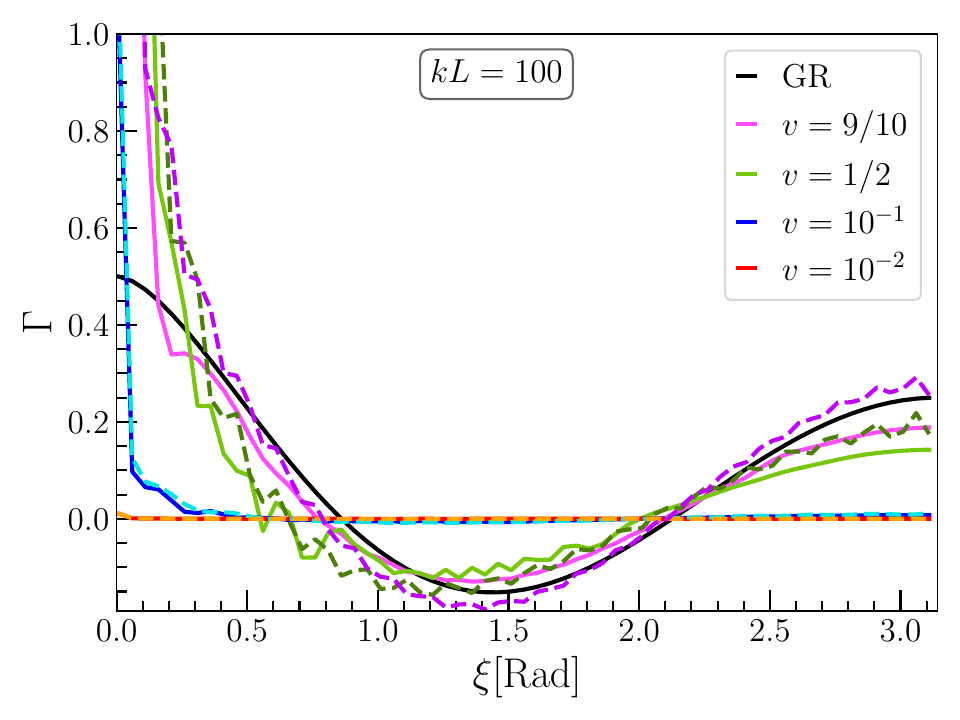}
\includegraphics[width=0.49\columnwidth]{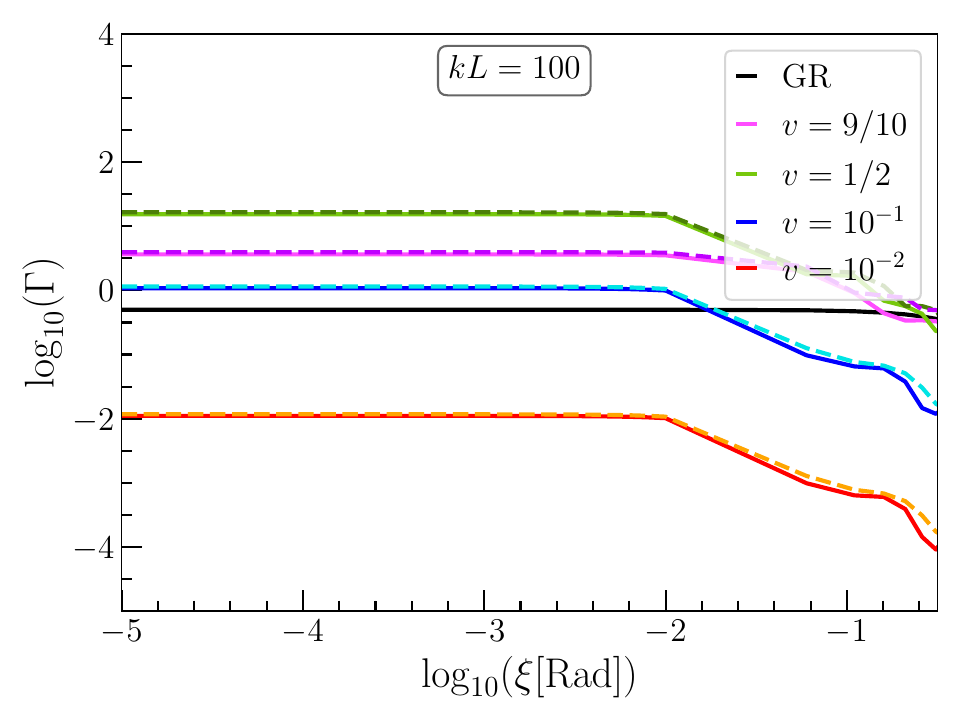}
\includegraphics[width=0.49\columnwidth]{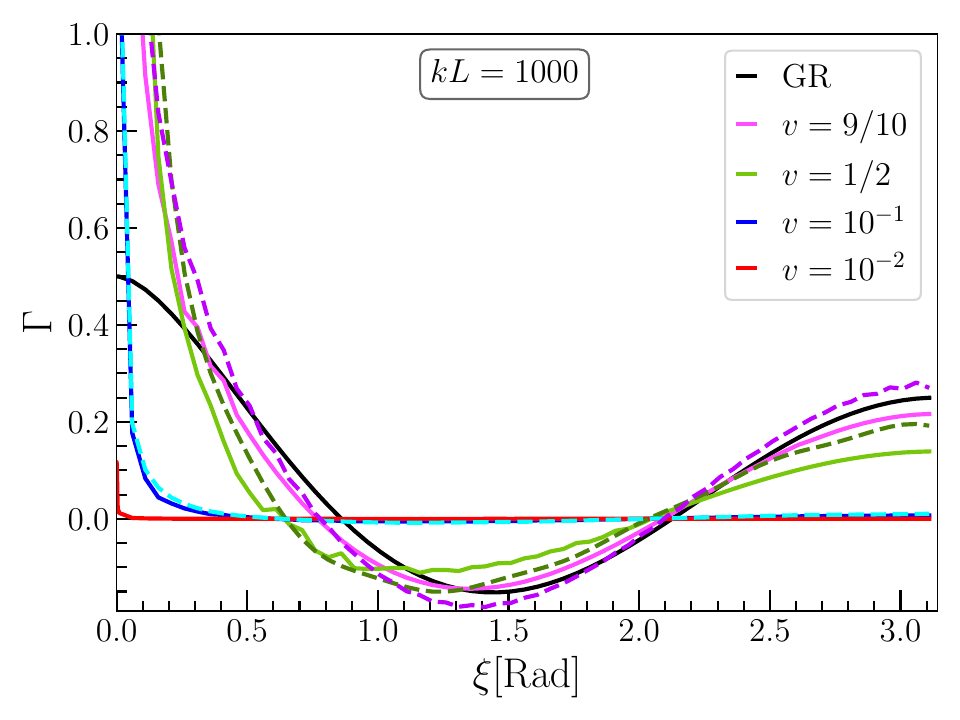}
\includegraphics[width=0.49\columnwidth]{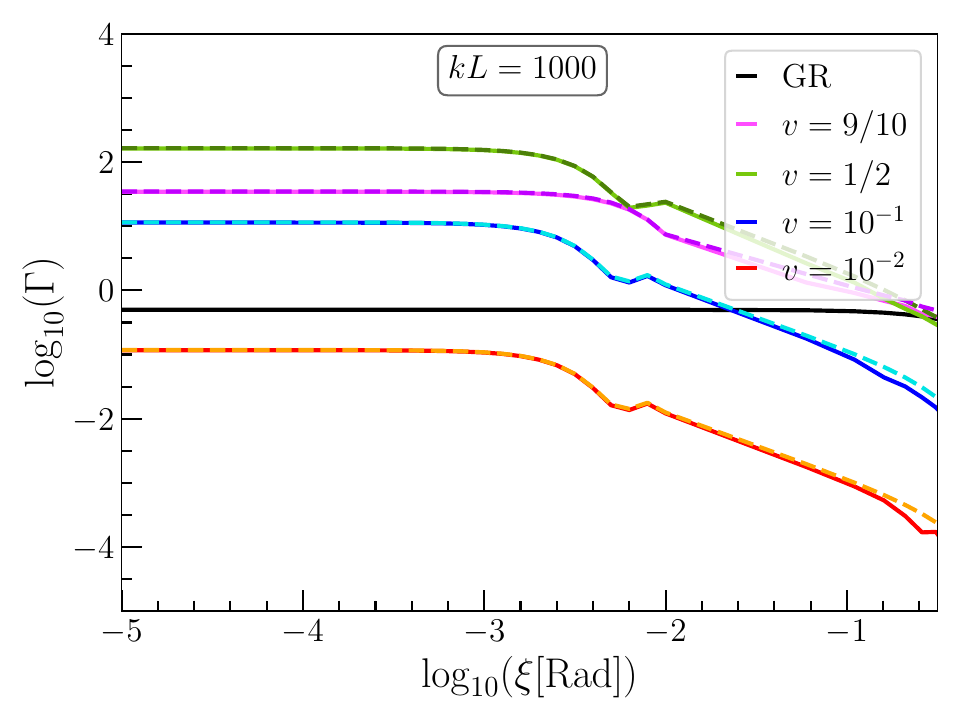}
\caption{ORF \eqref{eq:HDharmonic} for $v<1$ with fixed $kL=100$ (top panels) and $kL=1000$ (bottom panels). We show the numerical results of the ORF in solid lines and the analytical approximation given by Eq.~\eqref{eq:HDharmonic2} with nearby dashed lines with similar colors. The black line shows the result of GR ($v=1$), while the magenta, green, blue, and red lines, respectively, show the numerical ORF for $v=\{9/10,1/2,10^{-1},10^{-2}\}$. Purple, dark green, cyan, and orange dashed lines, respectively, show the analytical approximation for the same values of $v$. In the left panel, we show the full ORF. In the right panel, we zoom in on the small angle limit and use a logarithmic scale. Note how the numerical results and the analytical approximation agree well for $\xi\ll1$. The asymptotic value for $\xi\to0$ coincides with Eq.~\eqref{eq:gammaestimate}. We also checked that the analytical curve for $kL=1000$ matches the resummed approximation of Refs.~\cite{Liang:2023ary,Cordes:2024oem} in the range of the left plot.}\label{fig:HD} 
\end{figure}

We show in Fig.~\eqref{fig:HD} the ORF \eqref{eq:HDharmonic} for $kL=100$ (top panels) and $kL=1000$ (bottom panels) and $v=\{9/10,1/2,10^{-1},10^{-2}\}$. We also include the GR case ($v=1$) for comparison. In solid lines, we show the numerical result. The nearby dashed lines show the analytical approximation given by Eq.~\eqref{eq:HDharmonic2} with $c_\ell$ from Eq.~\eqref{eq:clnice3} together with Eq.~\eqref{eq:Iinfinity}. First, look at the top and bottom panels on the right. Note how the analytical approximation agrees well with the numerical one in the small $\xi$ limit. We also see that the ORF approaches a constant for values of $\xi$ below $\xi\sim 2/\ell_{\rm cut}$. We estimated such value of $\xi$ by requiring that the next order correction in a Taylor series expansion of the Legendre polynomials for $\xi\ll1$ becomes of ${\cal O}(1)$. This estimate yields $\log_{10}\xi\approx -1.7$ and $\log_{10}\xi\approx -2.6$ respectively for $kL=100$ and $kL=1000$, which is a reasonable estimate compared with the right panel of Fig.~\eqref{fig:HD}. We also see that the asymptotic value for $\xi\to0$ matches the analytical approximation given by Eq.~\eqref{eq:gammaestimate} in the large $kL$ limit.

Now, looking at the left top and bottom panels of Fig.~\eqref{fig:HD} we see that the numerical and analytical approximation are similar, though they do not overlap as much as in the small $\xi$ regime. We see that for $kL=100$, there are more wiggles than $kL=1000$ in the numerical and analytical cases due to finite distance effects. Increasing the value of $kL$ smooths the curve. Besides the wiggles, we still see some differences between the numerical and analytical results, even for $kL=1000$. There it is likely that some component is due to numerical errors as the ${\cal I}_\ell$ \eqref{eq:mainintegral} is the integral of highly oscillatory functions. Nevertheless, it is plausible that finite distance effects remain even for relatively large $kL$. From Fig.~\ref{fig:al}, it is clear that the analytical approximation given by Eq.~\eqref{eq:alnice} does not capture the behavior of $a_\ell$ near $\ell_{\rm cut}$. It would be interesting to carry out a more detailed and precise numerical study and, perhaps, refine the analytical approximations to the ORF for $v<1$.

\section{Conclusions \label{sec:conclusion}}

The ORF in PTAs provides a complementary way to test gravity by measuring the phase velocity $v$ of the GWs. In general, modified gravity predicts a phase velocity with $v\neq 1$, such as in scalar-tensor theories (see Ref.~\cite{Kobayashi:2019hrl} for a review), and possible with a $k$ dependence, as in massive gravity \cite{deRham:2019ctd}. Thus, departures of $v=1$ in the ORF would hint at modifications of gravity.

Motivated by recent studies \cite{Liang:2023ary,Liang:2024mex}, that encountered a divergent ORF in the small angle limit when $kL\to\infty$ for the $v<1$ case, we investigated finite distance effects on the ORF. To do that, we reformulated the calculation of the harmonic coefficients $c_\ell$ in terms of a definite integral of spherical Bessel functions given by Eq.~\eqref{eq:mainintegral}. The resulting formulation is similar to the equations in the TAM formalism \cite{Dai:2012bc,Qin:2018yhy,Qin:2020hfy,Liu:2022skj,Bernardo:2022rif,Bernardo:2022xzl,Bernardo:2022vlj,Bernardo:2023pwt} (see also App.~\ref{app:TAM}). As we argued at the beginning of Sec.~\ref{sec:calculations}, the approximation derived by Refs.~\cite{Liang:2023ary,Liang:2024mex} is valid up to $\ell_{\rm cut}\sim \sqrt{1-v^2}\,kL$ (see Eq.~\eqref{eq:lcut}). For $\ell\gtrsim\ell_{\rm cut}$ the integral (and $c_\ell$) decays exponentially, roughly as $\left({\ell}/{kL}\right)^{3/4}\times e^{-\ell\sqrt{{\ell}/{kL}}}$, see Eq.~\eqref{eq:Ihankel} and the discussion around it. For $\ell\gg kL$ it then decays as ${e^\ell}/{\ell}\times\left({kL}/{\ell}\right)^{\ell+1}$, see Eq.~\eqref{eq:Ileadingorder}.

The physical interpretation of our results, based on the nice physical explanations of Ref.~\cite{Romano:2023zhb}, is as follows. First, the factor $kL$ is the ratio of the “arm-length” (the distance to the pulsar) to the wavelength of the GW. Namely, $kL$ tells us how many GW oscillations undergoes the pulse from the pulsar until Earth. More precisely, the factor $kL$ enters in the redshift transfer transfer function (roughly the term inside parenthesis in Eq.~\eqref{eq:cloriginal}) in the antenna pattern of PTAs, yielding an angular pattern where points with vanishing redshift are separated by an angle $v+\cos\xi=2n/kL$ where $n\in \mathbb{Z}^+$ (see also Fig.~5 and 6 of Ref.~\cite{Romano:2023zhb} and Fig.~\ref{fig:antenna}). Thus, the typical angular separation of these features is $\xi\sim 1/kL$. Then, in the spherical harmonic decomposition, where the multipole number is roughly related to the inverse angular separation (i.e. $\ell\sim \pi/\xi$), it is expected that multipole numbers with $\ell\gtrsim kL$ contain no more information of the antenna response and their coefficients are, therefore, suppressed. In this work, we estimated more precisely the value of the cut-off to be at around $\ell_{\rm cut}\sim \sqrt{1-v^2}kL$.

Focusing on the small angle limit of the ORF, we found that using the large $\ell$ approximation for $a_{\ell}$ \eqref{eq:alnice}, first derived by Refs.~\cite{Liang:2023ary,Liang:2024mex}, up to the cut-off $\ell_{\rm cut}$ \eqref{eq:lcut} exactly recovers the value of the autocorrelation derived in Ref.~\cite{Cordes:2024oem}. We also checked our analytical estimates for the ORF numerically for $kL=1000$ and found a good agreement in the small angle regime. We thus showed that the ORF for $v<1$ is finite and that it correctly approaches the limiting value for $\xi\to 0$. Outside the small $\xi$ regime, we find that the numerical and analytical approximations are similar, although they do not match as well as in the small $\xi$ limit. This difference could be partly due to numerical errors, but it is possible that finite distance effects on the ORF will remain for relatively large values of $kL$. Basically, we find that the $a_\ell\approx {\rm constant}$ approximation \eqref{eq:alnice} for $\ell\gg 1$ is not good for values close to $\ell\sim \ell_{\rm cut}$. It is therefore important to carry out more precise numerical studies and perhaps refine the analytical templates of the ORF for $v<1$.

We also note that while we focused on the case $v<1$ our approach is also valid for any value of $v$. However, for $v\geq 1$ the numerical integration of Eq.~\eqref{eq:mainintegral} might be more challenging in the regime $\ell\ll kL$, with $kL\gg1$, as the oscillations from the exponential have a higher frequency and there are non-trivial cancellations. Nevertheless, for $v>1$ there are already analytical expressions for $c_\ell$ after dropping the exponential in \eqref{eq:cloriginal}, see, e.g., Refs.~\cite{Liang:2021bct,Wu:2023pbt,Wu:2023rib,Cordes:2024oem}.  Thus, our approach is most suitable for $v<1$ as it is sensitive to finite distance effects. For instance, we could numerically compute $c_\ell$ up to $\ell=1100$ for $kL=1000$, which is challenging to do in the formulation of Legendre polynomials \eqref{eq:cloriginal}. Interestingly, Ref.~\cite{Liang:2024mex} tested the value of $v$ using the ORF of CPTA data. They found that while not statistically significant there are two preferred values of $v=0.98$ and $v=1.04$ in the data. For $v<0.4$ though, the statistical significance dropped below $3\sigma$. It would be interesting to derive more precise templates with our formulation for $v\lesssim 1$ and test them against the data.

Lastly, we note that throughout the paper, we assumed equal distance pulsars for simplicity. It would be interesting to redo the calculations when the distance to the pulsars is different from each other, which is the more realistic situation. We leave this issue for future work, as it is out of the scope of this paper. We end by speculating that it may be possible to derive an exact formula for the coefficients $a_\ell$ without assuming the limit $kL\gg 1$. Looking at the exact expression for the autocorrelation given by Eq.~(23) of Ref.~\cite{Cordes:2024oem}, we notice that all terms in that formula can be expressed as an infinite sum of the product of spherical Bessel functions and Legendre polynomials (see, e.g., Secs.~6.10 and 10.6 of Ref.~\cite{NIST:DLMF}). It is thus plausible that $a_\ell$ may also be written as a series expansion in terms of such functions. In the worst-case scenario, it may only be possible in the exact $\xi\to 0$ limit. We leave this exercise for future work.

\section*{Acknowledgments} 
G.D. would like to thank B.~Allen, M.~Sasaki, K.~Schmitz and J.~Tr\"ankle for their helpful discussions. We also thank R.~C.~Bernardo and  Kin-Wang Ng for useful correspondence. G.D. is supported by the DFG under the Emmy-Noether program grant no. DO 2574/1-1, project number 496592360.

\appendix

\section{Comparison with the TAM formalism \label{app:TAM}}
In this appendix, we compare our formulas with those derived in the total angular momentum formalism \cite{Dai:2012bc}. There one defines, see e.g. \cite{Bernardo:2022rif},
\begin{align}\label{eq:gammaTAM}
\Gamma(kL,\xi)={\cal C}_{\rm TAM}\sum_{\ell=2}^\infty \frac{2\ell+1}{4\pi}C_\ell P_\ell(\cos\xi)\,,
\end{align}
where ${\cal C}_{\rm TAM}$ is a normalization constant,
\begin{align}\label{eq:ClTAM}
C_\ell = \frac{1}{\sqrt{\pi}}\left|{\cal J}_\ell(kL)\right|^2\,,
\end{align}
\begin{align}\label{eq:JellTAM}
{\cal J}_\ell(kL)=\sqrt{2}\pi i^\ell \sqrt{\frac{(\ell+2)!}{(\ell-2)!}}\,v \,I^{(2)}_\ell(kL,v)\,,
\end{align}
and we defined
\begin{align}\label{eq:Inell}
I^{(n)}_\ell(kL,v)=\int_0^{kL}dq \,\frac{j_\ell(q)}{q^n}e^{iqv}\,.
\end{align}

To show the equivalence with our formulation, we note that, after integration by parts, one has
\begin{align}\label{eq:I2ellone}
I^{(2)}_\ell=\frac{1}{\ell-1}\left(I_{\ell+1}^{(1)}-iv I_{\ell}^{(1)}+e^{ikLv}\frac{j_\ell(kL)}{kL}\right)\,.
\end{align}
Using the properties of the spherical Bessel functions and after more integration by parts, we further find that
\begin{align}
I^{(1)}_\ell=\frac{1}{\ell}\left(I_{\ell+1}^{(0)}-iv I_{\ell}^{(0)}+e^{ikLv}{j_\ell(kL)}\right)\,,
\end{align}
and
\begin{align}
I^{(1)}_{\ell+1}=\frac{1}{\ell+2}\left(I_{\ell}^{(0)}+iv I_{\ell+1}^{(0)}-e^{ikLv}{j_{\ell+1}(kL)}\right)\,,
\end{align}
We then conclude that Eq.~\eqref{eq:I2ellone} can be written in terms of Eq.~\eqref{eq:mainintegral} as
\begin{align}\label{eq:I2ell}
I^{(2)}_\ell(kL,v)=-\frac{1}{\ell(\ell+2)(\ell-1)}\Bigg(&\left(v^2(\ell+2)-\ell\right){\cal I}^*_{\ell}(kL,v)-2iv {\cal I}_{\ell}^*(kL,v)\\&-e^{ikLv}\left(j_\ell(kL)\frac{\ell-ivkL}{kL}-\ell j_{\ell+1}(kL)\right)\Bigg)\,,
\end{align}
having identified the functions defined in~\eqref{eq:Inell} and~\eqref{eq:mainintegral} via
    \begin{equation}
        \mathcal{I}^*_\ell(kL,v) = I^{(0)}_\ell(kL,v)\,.
    \end{equation}
Note that the terms inside the big parthensis in Eq.~\eqref{eq:I2ell} are the complex conjugate of the terms inside the brackets in Eq.~\eqref{eq:clnice}. It is then straightforward to check that both expressions \eqref{eq:gammaTAM} and \eqref{eq:HDharmonic} are equivalent up to a normalization factor, explicitly ${\cal C}_{\rm TAM}=\frac{\sqrt{\pi}}{4}{\cal C}$.

In passing, we provide a compact expression for the $kL\to\infty$ limit of Eq.~\eqref{eq:Inell}, which in Ref.~\cite{Qin:2020hfy} is given in terms of hypergeometric functions. This is given by
\begin{align}\label{eq:Inellinfinity}
I_\ell^{(n)}(kL\to\infty,v)=i^{l-n}\left|1-v^2\right|^{n/2}\left\{
\begin{aligned}
&\frac{\pi}{2} P_l^{-n}(v)+i \,Q_l^{-n}(v)&(v<1)\\
&i \,\Gamma[l-n+1]{\cal \mathbf Q}_l^{-n}(v)&(v>1)
\end{aligned}
\right.\,,
\end{align}
where $n<\ell+1$ and $P_l^{-n}(v)$ and $Q_l^{-n}(v)$ are the associated legendre polynomials for $v<1$. Note that ${\cal \mathbf Q}_l^{-n}(v)$ is defined by Eq.14.3.10 of Ref.~\cite{NIST:DLMF} for $v>1$, which follows Olver’s notation and it is real valued for real order and argument. More concretely, we have that
\begin{align}
{\cal \mathbf Q}^{-n}_{\ell}\left(v\right)=i^n \frac{Q^{-n}_{\ell}\left%
(v\right)}{\Gamma[\ell-n+1]}\,,
\end{align}
where $Q^{-n}_{\ell}\left(v\right)$ is the standard associated Legendre function for $v>1$.
Setting $n=0$ in Eq.~\eqref{eq:Inellinfinity}, one recovers Eq.~\eqref{eq:Iinfinity}.

\bibliography{refgwscalar.bib} 

\end{document}